**Personalised Medicine: Establishing predictive machine learning models for drug responses in patient derived cell culture**

Abbi Abdel-Rehim[1], Oghenejokpeme Orhobor[2], Gareth Griffiths[3], Larisa Soldatova[4], and Ross D. King[1,5,6,7]

[1]Department of Chemical Engineering and Biotechnology, University of Cambridge, Cambridge, UK.

[2]The National Institute of Agricultural Botany, Cambridge, UK.

[2]ValiRx Plc, Nottingham, UK.

[4]Department of Computing, University of London, London, UK.

[5]Department of Biology and Biological Engineering, Chalmers University of Technology, Gothenburg, Sweden.

[6]Department of Computer Science and Engineering, Chalmers University of Technology, Gothenburg, Sweden.

[7]The Alan Turing Institute, London, UK.

**Abstract**

The concept of personalised medicine in cancer therapy is becoming increasingly important. There already exist drugs administered specifically for patients with tumours presenting well-defined mutations. However, the field is still in its infancy, and personalised treatments are far from being standard of care. Personalised medicine is often associated with the utilisation of omics data. Yet, implementation of multi-omics data has proven difficult, due to the variety and scale of the information within the data, as well as the complexity behind the myriad of interactions taking place within the cell. An alternative approach to precision medicine is to employ a function-based profile of the cell. This involves screening a range of drugs against patient derived cells (or derivative organoids and xenograft models). Here we demonstrate a proof-of-concept, where a collection of drug screens against a highly diverse set of patient-derived cell lines, are leveraged to identify putative treatment options for a "new patient". We show that this methodology is highly efficient in ranking the drugs according to their activity towards the target cells. We argue that this approach offers great potential, as activities can be efficiently imputed from various subsets of the drug treated cell lines that do not necessarily originate from the same tissue type.

**Introduction**

A profound shift towards personalised cancer therapies is underway. This transformation is driven by the widespread recognition of the heterogeneity that characterises cancer, which is now considered to be a myriad of diseases united by uncontrolled growth. The diverse array of treatment strategies under development to treat cancer is a testament to the complexity of the disease.

Precision medicine is a transformative approach, tailoring treatments to the individual patients, while considering their unique genetic, environmental and lifestyle factors. This approach offers the potential for enhanced patient care and a reduction in the healthcare burden associated with these diseases. Nevertheless, a significant challenge still lies in the need for improved predictive models that can anticipate patient responses to specific treatments.

Recent breakthroughs in cultivating patient-derived cells in laboratory settings enables the establishment of in vitro models as valuable tools for advancing precision medicine. This methodology has the potential to inform on patient responses to treatment, deepening our understanding of underlying disease mechanisms, uncovering novel drug targets and ultimately enabling the development of more efficient treatment strategies.

Obtaining valuable insights from such models remains a challenge. While omics data has been used to varying degrees of success in this area, it often falls short of providing the versatility required for the diverse spectrum of available drugs and their mechanism of actions.

An alternative approach to omics is the utilisation of functional assays[1]. Studies have demonstrated that patient derived xenografts (PDX), can yield drug responses that correlate with the outcomes of patient responses in clinic[2]. However, the direct application of PDX models for precision medicine is often impractical due to significant cost, time, and resources necessary to generate results[3].

Addressing the relationship between ex-vivo responses and those observed in clinical settings remains a central and ongoing challenge in the realm of precision medicine. While studies correlating patient responses with those observed in patient derived cell culture (PDC) are limited, they are steadily increasing[4,5,6]. Notably, certain studies have demonstrated improved outcomes for blood cancer patients treated with drugs selected based on their ex-vivo performance, as opposed to standard treatments[7,8,9]. Addressing this challenge with vigour is paramount if we are to realize the full potential of this approach.

An essential step involves establishing an efficient methodology for precision medicine grounded in patient-derived cancer cell cultures. Growing cells in organoids provides a closer resemblance of the patient tumour than conventional tissue culture[10,11,12]. However, they often lack cells from the tumour microenvironment (TME) and can be costly to grow[4,13]. Whole-tumour cell culture is proposed as another alternative, which includes representation of cells from the TME at a lower cost than conventional organoids[4]. The large array of approved cancer drugs, coupled with numerous FDA approved available for consideration, presents a complex therapeutic landscape. While the straightforward approach of testing all drugs for all patients may seem ideal, it is prohibitively expensive and resource-intensive, particularly given the substantial number of cancer patients requiring screening.

An alternative, more practical approach is to harness the predictive capabilities of machine learning. Machine learning has been successfully applied in drug selection for a multitude of targets, as evidenced by a large number of publications in the field[14,15].

Various methods have been developed for the prediction of bioactivities in cell-lines and other cellular and molecular targets. The Quantitative Structure Activity Relationship (QSAR) approach, rooted in machine learning, is the most prevalent. It leverages molecular fingerprints based on chemical substructures to establish correlations with bioactivity in specific targets.

A different approach geared towards precision medicine makes use of omics data to predict the activity of drugs in individual patients. This method has demonstrated success for some drugs with clear correlations to specific dysregulated or dysfunctional genes[16,17]. However, a notable challenge remains, as no such method has yet successfully predicted the outcome of a diverse range of drugs across a varied patient population[18,19].

Here we introduce a promising approach that is rooted in bioactivity fingerprints and high throughput screening fingerprints (HTSFP), closely related to transformational machine learning[20,21,22]. This method relies on historical screening data as descriptors. These descriptors, much like bioactivity fingerprints, consists of bioactivities towards various targets. The key distinction is that a significant portion of these activities are actual measurements rather than mere predictions, enhancing the reliability and applicability of the methodology.

In our methodology, patient samples first undergo comprehensive screening of a drug library. A subset of this library is then selected as a probing panel. Subsequently, a new patient-derived cell line is screened only against this smaller panel. Machine learning is employed to learn relationships between drug responses found in historical samples and those in the new sample. This model trained on the responses from the drug panel, is then applied to predict drug responses across the entire library for the new cell line (Figure 1). Experimental validation can then be conducted on the top hits, and any confirmed hits represent potent candidates for a targeted drug cocktail tailored to the patient's cancer (Figure 2).

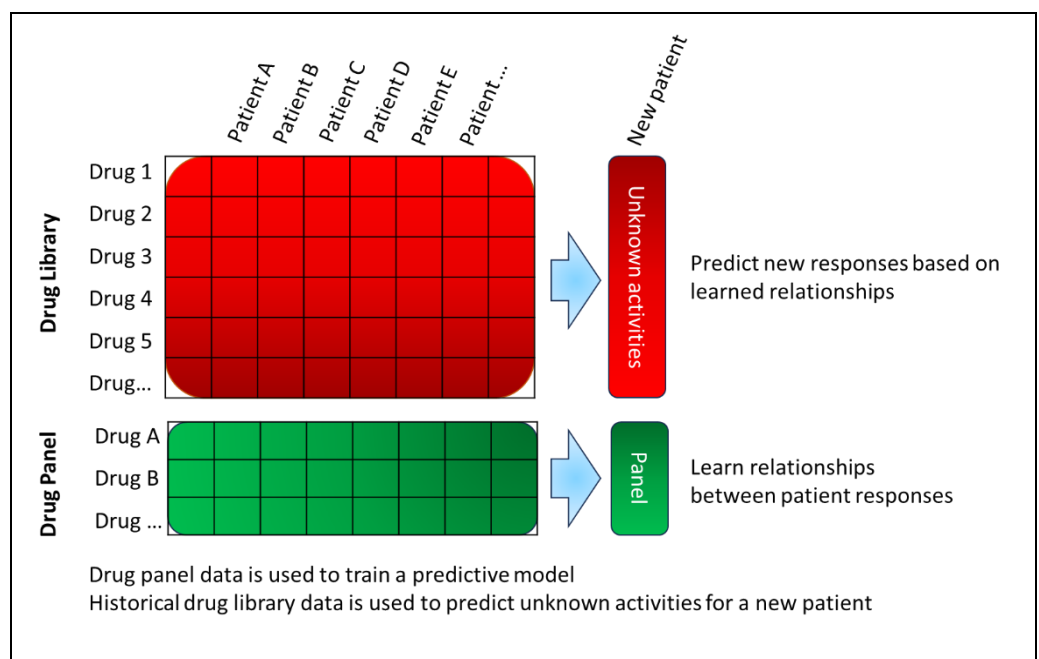

Figure 1. Principle of using HTS fingerprints for predicting activities in a new target cell line (patient derived). A machine learning model is generated using a drug panel where the new cell line has been screened against a limited set of compounds. These results are leveraged to predict activities from a larger library and acquire recommendations for potent drugs that can be validated in vitro.

In this hypothetical scenario, we are predicting drug sensitivities for patient derived cell lines through analysis of "historical" profiles of cell lines derived from other patients (Figure 2). Our findings demonstrate the significant efficiency of this approach in predicting drugs that exert a substantial impact on cancer cells. Moreover, we also establish the method's applicability in addressing the considerably more challenging task of identifying highly selective drugs for specific cancer cell lines. We provide additional evidence of the method's utility by applying it to a library of FDA-approved drugs. Finally, this methodology is also applied to a unique dataset with 24 patient derived tissue cultures screened against a limited panel of 35 FDA approved drugs shortly after the biopsies were performed.

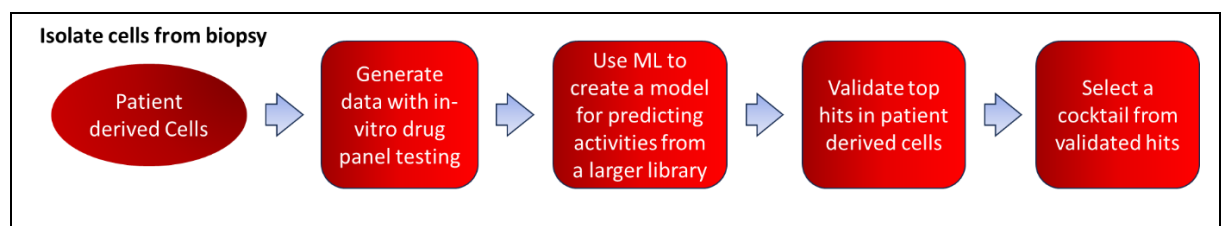

Figure 2. Workflow for exploiting patient derived cells from a patient's cancer to identify potent compounds from a large drug library. Patient derived cells are isolated from a biopsy and screened against a drug panel. The resultant data, together with historical screening data, is used to generate a model for predicting drugs from a larger library towards the new cell line. The top hits from the prediction are then validated and can inform a potent drug combination against the patient's cancer.

**Methods**

The RX dataset constituted tissue cultures derived from 24 patients with different cancer types screened against a drug library of 35 FDA approved drugs using a cell viability assay (Table S1).

Tissue collection and drug screening: All work was conducted in line with UK regulations after acquiring relevant approvals. Fresh samples of tumour specimen was acquired from hospitals and clinics in the UK. Upon arrival the samples were washed and processed according to a dissociation kit's instructions (Miltenyi Biotec , cat. 130-095-929) and gentleMACS dissociator (Miltenyi Biotec). After washing in PBS and a centrifugation, the cell pellet was resuspended in prewarmed medium and incubated for 48 hours at 37°C, 5% $CO_2$. Medium was changed to growth medium and cells were grown to approximately $1x10^6$ cells. Growth medium was changed every second day. Drugs were transferred to 384 well plates (Grenier cat. 781091) to result in a final concentration of 1 uM using a D300e digital dispenser (Tecan). Cell suspension was transferred to the wells with a density of 2000 cells per well in 50 uL growth medium. Cells were incubated with the drugs for 72 h at 37°C, 5% $CO_2$. Hoechst was used to stain the cells and acquire cell counts. The resultant values were used to calculate viability in relation to negative DMSO and positive Benzethonium chloride controls. The generated dataset was named RX.

Datasets: Four datasets were used in this study; the publicly available sets GDSC1, GDSC2 and PRISM[23,24], as well as the additional dataset (RX) described above.

Processing datasets: In the case of GDSC1 and GDSC2, we excluded cell lines with more than 20% missing drug measurements from the analysis. Similarly, drugs untested in over 20% of the remaining cell lines were also removed the study. For the larger PRISM dataset, we applied a more stringent threshold of 10% for cell line exclusion and 20% for drug exclusion. The resulting number of cell lines and drugs after this data pruning is summarised in table 1. For the RX dataset, given the limited size, we excluded cell lines with over 5 missing values, and drugs missing in over 5 cell lines.

Target values: For GDSC1 and GDSC2 2, we transformed target values to $\log(IC50)^{-1}$, with activity values above 6 (1 µM) considered active. In the case of PRISM, we utilized $\log2(\text{fold-change})^{-1}$ values as the target values, with activity values above 1.7 considered active. For RX, the dataset contained viability values normalised against DMSO controls, a value of 100 corresponded to 100% viability compared to DMSO treated cells. An active compound was indicated by the reduction of viability to <=30%.

Data completion: Since all the drug response datasets were incomplete, we considered two approaches for addressing this issue. The first involved filling the missing values in the matrices with zeros. Alternatively, we employed imputation using transformational machine learning (TML), an approach that essentially means generating a predictive model based on available values in the matrix to predict missing values. In this latter approach we used the sci-kit learn implementation GradientBoostingRegressor (n_estimators=50, max_depth=1). Initially, we excluded all test and validation samples (cell lines) ensure data integrity. The remaining cell lines designated as samples, and drugs were treated as features. Missing drug response values for each drug were predicted. For each drug under prediction, we generated a model based on all other drugs with fewer than 20 missing entries, any such entries were filled with zeros, and the missing values for the relevant drug were predicted and incorporated into a newly imputed "TML-matrix".

| Dataset | n-patients | n-drugs | n-drugs (targeted) |
|---|---|---|---|
| GDSC1 | 809 | 322 | 266 |
| GDSC2 | 707 | 165 | 137 |
| PRISM | 522 | 4515 | 4199 |
| RX | 24 | 35 | - |

Table 1. number of patients and drugs present in each dataset after data pruning according to specified thresholds. All drugs were used for the RX dataset as the drug library only constituted 35 compounds.

Regression: We conducted a comparison of regression models using the Sci-kit learn implementation RandomForestRegressor and GradientBoostingRegressor, both of which has shown strong performance in this kind of task[25,26,27,28]. We explored various configurations, including different number of trees for both learners, as well as adjustments to the max_depth parameter for GradientBoostingRegressor.

Evaluation metrics: To assess the performance of these regression models, we employed a range of evaluation metrics. These metrics included Pearson R, Root Mean Squared Error RMSE, standard deviations, and the precision of top n selections, presented as percentages. These metrics served to provide a comprehensive evaluation of the models predictive accuracy, robustness and suitability for the given task. Due to the smaller size of the RX dataset, leave-one-out cross-validation was used, and the metrics we chose were Spearman R, mean viability of the top 5 predicted hits, number of drugs predicted as hits as well as how many of these that were actual hits.

**Results**

To develop an efficient methodology for predicting drug responses in patient-derived cell lines we performed a few initial probing experiments. These provided the foundation for establishing parameters necessary to construct a viable prototype "recommender system", capable of predicting drug responses in unseen patient cell lines. We also wanted to explore the practical requirements for constructing a robust system. Two aspects were of particular interest, those are the number of historical patients necessary to train an efficient model, and the number of drugs needed for a potent probing panel. These calibration experiments were carried out using a dedicated validation set from the GDSC1 dataset, encompassing 81 patients, which accounts for 10% of all patients in the dataset.

Based on the outcomes of these experiments outlined in Appendix A, we used TML to fill in missing values in the training datasets (Table S2, Figure S1), we employed a random forest with 50 trees (default parameters) (Table S3), we used an initial "probing" drug panel of 30 selected drugs for all "unseen" cell lines (Table S4 and S5), and we included 100 randomly selected patients when training the models (Table S6). These choices constitute the foundation of our prototype system.

**Predictive performance in the GDSC1 dataset**

The prototype outlined above was applied to a dedicated test set, containing 81 patient derived cell lines. The performance was evaluated through various metrics. In addition to $R_{pearson}$, $R_{spearman}$ and RMSE, we also included four accuracy metrics.

We reported the percentage of accurate predictions within the top 10, 20 and 30 drugs, providing an indication of the predictive accuracy. For instance, if 7 out of the 10 predictions matched the actual top 10 drug responses, we reported a percentage of 70%. Additionally, we reported the hit-rate among the top-10 predictions, offering a direct insight into the number of these recommendations that accurately identified hits.

All the results are reported as a mean of five experiments, accompanied by the resultant standard deviations (Table 2). For the sake of comparison, the average hit-rate among the entire set of test-set drugs was approximately 17.8% for all drugs and 5.0% for "targeted" drugs (those active in less than 20% of the cell lines).

The prototype recommender system demonstrated excellent performance, with high correlations between predicted and actual drug activities for both the entire drug library and targeted drugs alike (Table 2). When considering the entire library, we found that the recommender system performed very well. On average 6.6 out of the top 10 predictions were correctly identified, with 15.26 and 22.65 accurate predictions for the top 20 and 30, respectively. Even when agnostic towards the exact ranking of the drugs and aiming for 10 recommended hits, the system consistently predicted almost only hits. In the most challenging task of predicting targeted drugs, the results remained high in terms of overall bioactivity ranking ($R_{pearson}$ = 0.781, $R_{spearman}$ = 0.791). When specifically aiming to rank identify the top 10, 20 and 30 drug cohorts, the system provided an average of 3.6, 10.5 and 17.6 accurate predictions, respectively. The hit-rate amongst the top 10 drugs was slightly higher, averaging at 4.3. It is important to note that, for the targeted drugs, 50% of all cell lines had 12 or less hits in total out of the 236 available for prediction, demanding a nearly perfect system to pick them out. If one only were to consider the 41 cell lines with more than 12 hits present, the hit-rate would increase to 6.1.

|  | $R_{pearson}$ | $R_{spearman}$ | RMSE | Top 10 | Top 20 | Top 30 | Hit-rate |
|---|---|---|---|---|---|---|---|
| Targeted drugs | 0.781/0.021 | 0.791/0.020 | 0.469/0.018 | 0.36/0.086 | 0.524/0.060 | 0.588/0.052 | 0.426/0.075 |
| All drugs | 0.885/0.008 | 0.865/0.010 | 0.507/0.017 | 0.66/0.063 | 0.763/0.027 | 0.755/0.028 | 0.978/0.011 |

Table 2: Predictive performance of the prototype drug recommender system on the GDSC1 dataset. Metrics reported are $R_{pearson}$, $R_{spearman}$, RMSE, and accuracy of predictions within the top 10, 20 and 30 drugs, as well as hit-rates among the top 10 recommendations. Results are presented as mean/stdev based on five experiments.

Looking closer at the performances across individual cell lines in terms of Spearman R coefficients, which indicates how well drugs are ranked in terms of their activity, we observed a minimum score of 0.76 for all drugs and 0.39 for targeted drugs (Figure S2). Notably, among the targeted drugs which are more difficult to predict, only 8 cell lines performed below 0.7 and only two fell below 0.65. These results illustrate the overall strong performance of the prototype recommender system while highlighting the increased complexity of predicting responses for targeted drugs. An additional experiment comparing our approach to the use of standard molecular fingerprints was also performed and comparative results shows that this approach is far superior on this dataset (cf. Table S9 and S10).

**Predictive performance in the GDSC2 dataset**
The parameters selected for the GDSC1 dataset were applied to the GDSC2 dataset, demonstrating consistent high performance across all drugs. In the GDSC2 dataset, the hit-rate for all test-set drugs was 13.2%, whereas for targeted drugs, it was 2.5%. These values are notably lower than those of the GDSC1 dataset. When considering all drugs, the hit rate amongst the top 10 recommendations was high with approximately 9/10 drugs being active on average (Table 3). However, the hit rate for targeted drugs in the top 10 predictions was 0.193, significantly lower than in GDSC1 (Table 2). This decline can again be attributed to the small number of hits available in the dataset, averaging 3.38 across all cell lines. Consequently, the systems performance is constrained by this limit, achieving a maximum average of 3.38 hits per 10 recommendations. With this limit in mind, the system's performance can be adjusted to 57% in accuracy when compared to the theoretical maximum. Furthermore, 21 cell lines lacked any hits at all, if these are removed, the average hit-rate goes up to 2.72, still standing at 57% of the maximum available hit-rate (now averaging 4.77). There were only 8 cell lines that had 10 or more hits, averaging at 13.25, and for these cell lines the average hit-rate was 6.6/10.

|  | $R_{pearson}$ | $R_{spearman}$ | RMSE | top 10 | top 20 | top 30 | Hit-rate |
|---|---|---|---|---|---|---|---|
| Targeted drugs | 0.775/0.026 | 0.767/0.025 | 0.460/0.021 | 0.543/0.084 | 0.680/0.046 | 0.744/0.035 | 0.193/0.042* |
| All drugs | 0.892/0.010 | 0.841/0.016 | 0.526/0.020 | 0.737/0.051 | 0.824/0.026 | 0.804/0.024 | 0.890/0.019 |

Table 3: Predictive performance of the prototype drug recommender system on the GDSC2 dataset. The reported metrics encompass $R_{pearson}$, $R_{spearman}$, RMSE, accuracy of predictions within the top 10, 20 and 30 drugs, and hit-rates among the top 10 recommendations. Results are presented as mean/stdev derived from five experiments. *Maximum possible hit-rate is 3.38.

**Predictive performance for an FDA-approved drug library**

The PRISM dataset differs significantly from previous dataset as it contains a larger library of FDA approved drugs. Due to the unique and diverse nature of this dataset, we again investigated the optimal size for the probing drug panel, as well as the impact of the number of cell-lines included in the training on model performance. We excluded the two compounds 'mg-132' and 'bortezomib' from our studies, as the authors of the PRISM study indicated their use as positive controls.

First, we explored the effect of the number of patients used in training the models. The experiments were performed using a drug panel containing 90 drugs (~2% of the dataset). Surprisingly, we found that reducing the number of patients from 418 (all patients) to 30 patients still retained a significant amount of information necessary efficient predictions (Table S7). Intriguingly, even when working with 10 patients, correlations remained relatively close to those of the larger patient subsets. However, for this minimal cohort, there was a marked drop in accuracy amongst the top-performing drugs, as well as hit-rates among the top 45 selections. Nonetheless, our study demonstrates that reducing the number of patients to approximately 10% of the entire dataset still yields strong predictions and a substantial number of hits in the recommendations. Based on our findings, we chose to proceed with 100 patients, as the performance was nearly identical to that of 200 and 418 patients.

Next, we investigated the impact of drug panel size on performance. We conducted experiments with drug panels consisting of 23, 45 and 90 drugs, representing approximately 0.5%, 1% and 2% of the dataset, respectively. We observed that there was a marked improvement with each size increase. However, considering that testing 90 drugs for every new patient could be quite a substantial, we recommend using a panel of 45 drugs (or less). This smaller panel still yielded competitive results and a substantial number of active compounds active against the cell lines, ~30 hits present in 45 recommendations (Table S8).

The calibration work above was performed using a validation set. Subsequently, we tested the same panel sizes using an independent test-set of 52 patients and observed very similar performance (Table 4). It became evident that the hit-rates (amongst the top 45 predicted drugs) corresponded to increases of approximately 1000-1600% compared to screening the entire library (5.2% hit-rate). This gives rise to 45 recommendations containing approximately 25-38 active compounds on average depending on the chosen panel size.

| Panel size | $R_{pearson}$ | $R_{spearman}$ | RMSE | top 0.5% | top 1% | top 2% | Hit-rate (top 45) |
|---|---|---|---|---|---|---|---|
| 23 | 0.530/0.045 | 0.394/0.035 | 0.776/0.012 | 0.067/0.036 | 0.169/0.059 | 0.368/0.087 | 0.559/0.103 |
| 45 | 0.609/0.024 | 0.444/0.022 | 0.742/0.009 | 0.083/0.046 | 0.215/0.070 | 0.437/0.078 | 0.662/0.099 |
| 90 | 0.687/0.019 | 0.473/0.015 | 0.684/0.010 | 0.159/0.064 | 0.345/0.074 | 0.567/0.054 | 0.845/0.063 |

Table 4. Impact of the different panel sizes on predictive performance using an independent test-set of 52 PDCs. The reported metrics includes $R_{pearson}$, $R_{spearman}$, RMSE, accuracy of predictions within the top 10, 20 and 30 drugs, and hit-rates among the top 10 recommendations. Results are presented as mean/stdev derived from five experiments.

**Predictive performance for RX dataset**

The RX dataset was acquired from a range of tumour types. The drug screens were conducted on cultured biopsies of cancer patients shortly after their arrival at a laboratory. Due to the limited number of drugs and cells in the dataset resulting from our stringent inclusion criteria (see methods), we employed leave-one-out cross-validation for this dataset. This involved predicting the viability of patient-derived cells when treated with a drug, based on their resultant viabilities when treated with the remaining drugs in the dataset as well as those of other patients. In this scenario, the unknown activities illustrated in Figure 1 would pertain to a single drug, while the remaining drugs would be part of the drug panel.

Approximately 30 different drugs were tested against each cell line, the average resultant viability across cell lines for this library was 85.78% (Table 5). In contrast, when selecting the five drugs predicted to be most effective towards each cell lines, the mean viability of the cancerous cells dropped to 32.16%. Using the threshold of <30% viability as an indication for a hit, the average number of hits available per cell line was 4.57. On average, the methodology recommended 5 drugs for testing and 3.74 of these were identified as hits. Three of the cell lines never experienced a viability decrease of <30% and were excluded from the hit analysis. The average hit rate across all cell lines was 72.6%, and our approach managed to identify an average of 68.8% of all available hits across all cell lines. For three cell lines, the hit rate was 0%, and in each case, the total hits amongst the 30+ compounds were low, at 1 or 2 hits. However, the methodology always managed to rank drugs in the library so that the top 5 predicted drugs resulted in a significantly lower viability than that of the library. Complete results can be found in Table S11.

| $R_{spearman}$ | Library (viability) | Top 5 ranked (viability) | n hits available | n hits predicted | n hits identified | drugs in panel |
|---|---|---|---|---|---|---|
| 0.62/0.02 | 85.78 | 32.16/2.83 | 4.57 | 5.06/0.28 | 3.77/0.12 | 33.04 |

Table 5. Performance on RX dataset. For the 24 cell lines, the mean Spearman rank is given for the entire library of 30+ drugs. The average viability resulting from the drug library is reported, along with that found when selecting the top 5 predicted drugs. Defined as reducing cell viability to less than thirty percent, the average number of hits available per cell line is shown, as well as the average number of drugs predicted to be hits, and the average number of these correctly identified as such. The average number of drugs tested per cell line is found in the last column "drugs in panel". The results from five independent experiments are shown as mean/standard deviation.

**Discussion**

The future of cancer therapy is moving towards a personalised approach, aiming to align treatment strategies with the needs of the individual patient. The approach presented in this work seeks to leverage weaknesses of the tumour while minimising harm to the healthy body. Current efforts in this area are focused on linking specific treatment options to genetic markers, but the choices of such personalised drugs are relatively few, and this approach only benefits ~10% of all patients[29]. The challenge only becomes greater when selecting medications that lacks a direct link to specific gene defects.

Efforts to connect therapy choices to genetic makeup have encountered limitations in their success[18,19]. Most promising studies have concentrated on a select few drugs rather than a larger library or a panel of currently available drugs. These limitations are rooted in the complexity of the cell and its dynamic molecular regulation at multiple levels, rendering it challenging to predict drug outcomes based on such complex and interconnected information.

In this context, the methodology presented in this work set out to circumvent this complexity by shifting the focus away from a static depiction of a cell's state and its potential response. Instead, it explores the cell's ability to respond to a diverse drug panel, aiming to understand its interaction with drugs outside this panel.

This study delves into drug response profiles as predictors in precision medicine, seeking to establish fundamental principles for constructing robust machine learning models for recommending drugs effective against patient derived cancer cells. The correlations between number of patient derived cells, drug panel size and model performance provide valuable insights into the key parameters influencing model performance. The experiments indicate that the optimal range for achieving efficient predictions falls within 100-200 patients in the training data, with incremental gains in performance beyond this (Tables S5 and S6). As for the drug panel, depending on the type of library one wants to extract recommendations from, one will have to use varying numbers of drugs in the probing panel. For instance, the experiments demonstrate that even a probing panel as small as 10 drugs can yield high performance in predicting activities in the GDSC1 dataset consisting of ~300 anti-cancer compounds (Table 2). The experiments also showed that a small probing panel of 23 drugs resulted in an average of 25 hits amongst 45 selected from an FDA-library of ~4500 compounds (Table 4).

Notably, this methodology exhibits a remarkable level of agnosticism towards the specific type of cancer tissue from which the cells are derived. This finding is further underscored when investigating the composition of the ten most informative cancer cells (according to their tissue types) when predicting example cell lines from different tissues of origin (Figure S3, Figure 3). Only haematopoetic and lymphoid tissue cancer appears to prefer similar cancer types for efficient prediction. In addition, we found that the specific selection of drugs that is included in the panel, as well as the size of the panel, have a profound impact on how patients cluster (Figure S4), consistent with previous studies[30]. The clustering becomes more robust when increasing the size of the panels, suggesting that overall, specific cell lines tend to be similar in their general response (hints at the concept of digital twins in precision medicine).

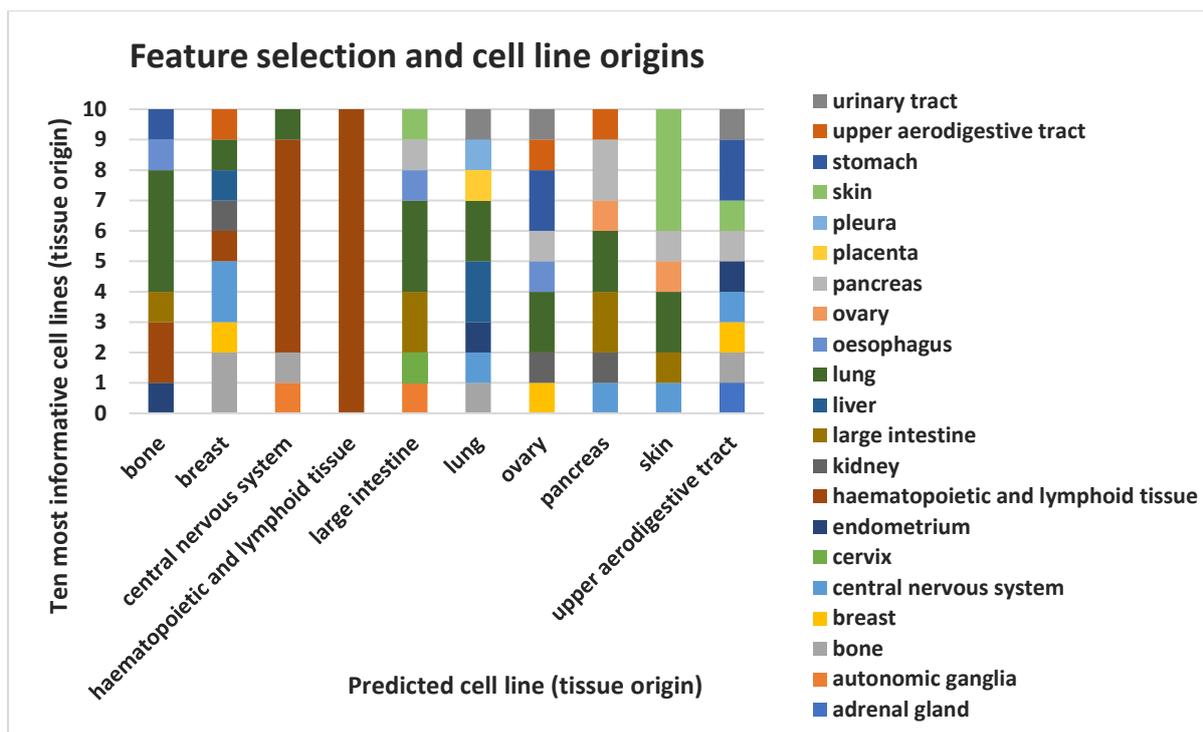

*Figure 3. Feature selection results. For each of 10 cell lines being predicted (defined by their tissue of origin), the most influential cell-lines were identified, and their respective tissue of origin reported. 50% of the drug library was used in the probing panel.*

Lastly, this methodology demonstrated strong performance on the RX dataset, which includes tissue cultures exposed to drugs shortly after biopsy. This serves as a crucial validation, highlighting the approach's potency with clinical samples, even when the drug library is limited and the number of patients is small. In nearly every case where hits were present, at least one was recommended. Only in 3 out of 21 cell lines, where there were only 1 or 2 hits, did the approach fail to identify one. However, the methodology was still successful; for every cell line, the mean viability of the top 5 hits was significantly lower than that of the entire library, suggesting that the methodology correctly identified effective compounds from this library (Table S11).

Clinicians have already seen promising outcomes using PDC data to guide treatment decisions, particularly regarding drug resistance. Our methodology offers a streamlined and cost-effective solution, requiring only a fraction of a drug library for initial screening and capable of delivering a tailored selection of effective drugs within a week, revolutionizing treatment timelines. Recognizing the diverse drug susceptibilities of cancer cell lines, our method supports the formulation of strategic drug cocktails to optimize treatment, minimize redundancy, and prevent resistance.

It is important to note that the methodology presented in this work is based on data from cell-based assays. This means that challenges may arise when studying certain classes of drugs such as immunomodulators. Acknowledging the growing significance of modulators influencing the microenvironment and immune system, future iterations of this work must address these nuances. Strategic drug cocktails, including modulators and various therapeutic agents, are likely to be at the core of future precision medicine. Our methodology stands poised to contribute to these efforts by facilitating potent and tailored therapies.

**Acknowledgements**

Code Availability: Code can be found in a dedicated github repository abbiAR/DrugResponsePrediction (github.com).

Data Availability: RX dataset can be found in Supplementary Table 1.

None of the above work would have been possible without the availability of cancer derived cell lines. We'd like to express gratitude to all patients who have made this, as well as many other studies contributing to the understanding and treatment of cancer, a possibility.

This work has been supported by the UK Engineering and Physical Sciences Research Council (EPSRC) [EP/R022925/2, EP/W004801/1 and EP/X032418/1].